\documentclass[10pt,aps,notitlepage,nobibnotes,
prl,floatfix,twocolumn,showpacs,citeautoscript,
preprintnumbers,superscriptaddress]{revtex4-1} 

\usepackage[english]{babel} 	
\usepackage{amsmath}        	
\usepackage{amssymb}        	
\usepackage{bm}					
\usepackage{bbm}            	
\usepackage{etoolbox} 			
\usepackage{graphicx}       
\usepackage{graphics}       
\DeclareGraphicsExtensions{.pdf,.jpg} 
\usepackage{hyperref}
\hypersetup{colorlinks=true, pdfstartview=FitV, 
linkcolor=blue, citecolor=blue, urlcolor=blue}

\makeatletter         
\newcounter{firstbib} 
\apptocmd{\thebibliography}{ \setcounter{NAT@ctr}{\value{firstbib}} }{}{} 
\AtEndEnvironment{thebibliography}{ \setcounter{firstbib}{\value{NAT@ctr}}}
\makeatother

\newcommand{\Equation}[2]{\begin{equation}\label{#1}#2\end{equation}}

\newcommand{\SubAlign}[2]{\begin{subequations}\label{#1}\begin{align}#2\end{align}\end{subequations}}

\newcommand{\bs}{\boldsymbol}
\newcommand{\Figref}[1]{Fig.~\ref{#1}}
\newcommand{\Eqref}[1]{\eqref{#1}}

\newcommand{\groupU}[1]{\mathrm{U}(#1)}   
\newcommand{\groupSU}[1]{\mathrm{SU}(#1)} 
\newcommand{\groupZ}[1]{\mathbb{Z}_{#1}} 			
\newcommand{\groupS}[1]{\mathbb{S}^{#1}} 			
\newcommand{\Exp}[1]{\text{e}^{#1}}
\renewcommand\Re{\mathrm{Re}}
\renewcommand\Im{\mathrm{Im}}
\newcommand{\Grad}{{\bs\nabla}}

\newcommand{\Curl}{{\bs\nabla}\times}

\newcommand{\A}{{\bs A}}
\newcommand{\B}{{\bs B}}
\newcommand{\D}{{\bs D}}
\newcommand{\E}{\mathcal{E}}
\newcommand{\Q}{{\cal Q}}
\newcommand{\J}{{\bs J}}

\usepackage[singlelinecheck=false, justification=RaggedRight, format=plain]{caption}
\DeclareCaptionLabelSeparator{bar}{ $\boldsymbol{\mid}$ }
\begin{document}
\title{Kelvin knots in superconducting state}

\author{Filipp~N.~Rybakov}  
\affiliation{Department of Physics, KTH-Royal Institute of Technology, 
Stockholm, SE-10691 Sweden} 
\author{Julien~Garaud} 
\affiliation{
Institut Denis-Poisson CNRS/UMR 7013, \\
Universit\'e de Tours - Universit\'e d'Orl\'eans, 
Parc de Grandmont, 37200 Tours, France}
\author{Egor~Babaev} 
\affiliation{Department of Physics, KTH-Royal Institute of Technology, 
Stockholm, SE-10691 Sweden}

\date{\today}
\maketitle

{\bf 
The failed ``vortex-atoms" theory of matter by Kelvin and Tait 
\cite{Thomson:67,Tait:98} had a profound impact on mathematics and physics. 
Building on the understanding of vorticity by Helmholtz, and observing stability 
of smoke rings, they hypothesised  that elementary particles (at that time atoms)
are indestructible knotted vortices in luminiferous aether: the hypothetical 
ideal fluid filling the universe.
The vortex-atoms theory identified chemical elements 
as topologically different vortex knots, and matter was interpreted as bound 
states of these knotted vortices.
This work initiated the field of knot theory in mathematics. It also  
influenced modern physics, where a close although incomplete analogy exists 
with the theory of superfluidity, which started with Onsager's and Feynman's 
introduction of quantum vortices \cite{Onsager:49,Feynman:55}. Indeed many 
macroscopic properties of superconductors and superfluids are indeed determined 
by vortex lines forming different  ``aggregate states", such as vortex crystals 
and liquids.  
While crucial importance of knots was understood for many physical systems 
in the recent years, there is no known physical realization of the central 
element of Kelvin theory: the stable particle-like vortex knot. Indeed, 
vortex loops and knots in superfluids and ordinary superconductors form as 
dynamical excitations and are unstable by Derrick theorem \cite{Derrick:64}. 
This instability in fact dictates many of the universal macroscopic properties 
of superfluids.
Here we show that there are superconducting states with principally different 
properties of the vorticity: where vortex knots are intrinsically stable.   
We demonstrate that such features should be realised near certain critical 
points, where the hydro-magneto-statics of superconducting states yields 
stables vortex knots which behave similar to those envisaged in Kelvin and 
Tait's theory of vortex-atoms in luminiferous aether.
}

Kelvin's theory was falsified, when Michelson and Morley's experiment ruled 
out the existence of aether. Yet the principle to associate vortices in some 
underlying field with ``elementary particles" re-emerged in two important 
concepts in modern condensed matter physics: the particle-vortex duality and 
the interpretation of collective vortex states as ``vortex matter'', most 
notably in superfluidity and superconductivity. These analogies follow 
developments of three paradigm-shifting concepts introduced in Onsager's work 
on superfluids \cite{Onsager:49}. 
First concept was the observation that superfluid velocity circulation is 
quantized, hence vortices carry a quantized topological charge. 
Second observation was that rotation of a superfluid results in the formation 
of a lattice or a liquid of quantum vortices, i.e. the vortex matter realisation 
of crystals and liquids. 
The third crucial concept is that vortex matter controls many of the key 
responses of superfluids. For example, the superfluid to normal state 
phase transition is a thermal generation and proliferation of vortex loops 
and knots \cite{Onsager:49}. Subsequently this theory was put on firm 
theoretical grounds by Feynman \cite{Feynman:55}. Superconducting phase 
transition was demonstrated to be driven likewise by proliferation of vortex 
loops \cite{Dasgupta.Halperin:81}.
A remarkable analogy with Kelvin's theory resides in the Berezinskii, Kosterlitz 
and Thouless theory of two-dimensional superfluids, where vortices with opposite 
circulations are mapped onto particles and antiparticles.
In three dimensions the thermal and quench responses, and turbulent states 
are collective states of vortex loops and knots. However, the crucial 
difference with the Kelvin's theory is that vortex loops and knots are 
intrinsically unstable, as follows from Derrick's theorem \cite{Derrick:64}. 
This implies that an excited system forms vortex loops and knots which, 
however, tend to collapse as the kinetic energy of the superflow always 
decreases for a smaller loops or knots.

Research on models supporting stable knotted solitons has been of great 
interest in mathematics and physics after stability of these objects was 
found in the so-called Skyrme-Faddeev model \cite{Faddeev:75,
Gladikowski.Hellmund:97, Faddeev.Niemi:97} (for a review, see 
\cite{Radu.Volkov:08}).
It was further observed that there exist a formal relation between 
Skyrme-Faddeev's model and ostensibly unrelated, Ginzburg-Landau theories 
for multicomponent superconductors \cite{Babaev.Faddeev.ea:02,Babaev:02b}. 
Namely, two-component Ginzburg-Landau models can be mapped onto a 
Skyrme-Faddeev model coupled to an additional massive vector field.
This observation motivated the conjecture that multicomponent superconductors
may support stable knots. Detailed numerical studies however did not found 
stability \cite{Jaykka.Hietarinta.ea:08}. The reasons for the instability 
were subsequently discussed both using physical estimates \cite{Babaev:09}, 
and formal mathematical approach \cite{Jaykka.Speight:11}.
Despite different analytical arguments in favour of the (meta)stability 
\cite{Babaev:09,Gorsky.Shifman.ea:13}, and findings of the stability of 
knots in mathematical generalisations of Skyrme-Faddeev model coupled to gauge 
fields \cite{Ward:02,Jaykka.Speight:11}, the prevalent opinion today is that 
knotted vortices are unstable in theories of superconductivity as in superfluids.

We demonstrate in this paper, that stable vortex knots exist in two-component 
superconducting states in a certain parameter range. 
Many of the superconducting states of interest today have multiple components, 
for various reasons: e.g. spin-triplet pairing, or nematic states  (for recent 
examples see e.g. \cite{Wang.Cho.ea:16,Wang.Fu:17}), or coexistence of 
superconductivity of electrons and nucleons \cite{Sjoberg:76,Babaev.Sudbo.ea:04,
Babaev.Ashcroft:07,Jones:06}.
A generic feature of multicomponent superconductors and superfluids is the 
existence intercomponent current-current interaction, also known as the 
Andreev-Bashkin effect \cite{Andreev.Bashkin:75,Svistunov.Babaev.ea}. Namely, 
in superfluid mixtures of two components (labelled ``1'' and ``2''), because 
of the intercomponent interaction between particles, the current of a given 
component ${\bs j}_{1,2}$ generically depends on the superfluid velocities 
${\bs v}_{1,2}$ of both, as follows:
\Equation{Eq:Andreev-Bashkin}{
{\bs j}_1 = \rho_{11}{\bs v}_1+\rho_{12}{\bs v}_2\,,~~~\text{and}~~~
{\bs j}_2 = \rho_{22}{\bs v}_2+\rho_{21}{\bs v}_1\,.
}
There, the coefficients $\rho_{12}$ and $\rho_{21}$ determine the fraction of 
the density of one of the superfluid component carried by the superfluid velocity 
of the other: i.e. the intercomponent drag. The drag coefficients $\rho_{12}$ 
and $\rho_{21}$ can be very large, for example, in spin-triplet superconductors 
and superfluids \cite{Leggett:75}, Fermi-liquids mixtures \cite{Sjoberg:76}
or strongly correlated systems \cite{Kuklov.Svistunov:03,Sellin.Babaev:18}.

Two-component superconductors are described by a doublet
$\Psi=(\psi_1,\psi_2)^{\mathrm{T}}$, of complex fields 
$\psi_a=|\psi_a|\Exp{i\varphi_a}$ (with $a=1,2$), whose squared moduli 
$|\psi_a|^2$ represent the density of individual superconducting components. 
Each of the components is coupled, via the gauge derivative $\D=\Grad+ig\A$, 
to the vector potential $\A$ of the magnetic field $\B=\Curl\A$. 
Such a system is described by the Ginzburg-Landau free energy $E=\int \E d{\bf r}$, 
whose density reads as:
 \SubAlign{Eq:FreeEnergy}{
 \E&= \frac{\B^2}{2}+\sum_{a=1,2}\frac{\gamma_a}{2}|\D\psi_a|^2   	
  	 +\sum_{a,b=1,2}\frac{\mu_{ab}}{2} {\bs J}_a\cdot{\bs J}_b  	
  \label{Eq:FreeEnergy:1} \\
  &~~~~~~
   +{\nu}\left(\Psi^\dagger\Psi-1\right)^2
   +V[\Psi,\Psi^\dagger]
  		\label{Eq:FreeEnergy:2}\,. \\
  &\text{where}~~{\bs J}_a=\Im(\psi_a^*\D\psi_a)=|\psi_a|^2(\Grad\varphi_a+g\A) \,.
\nonumber
}
The terms $\mu_{12}=\mu_{21}$ of the current coupling matrix $\hat{\mu}$, 
describe the intercomponent drag \cite{Andreev.Bashkin:75,Leggett:75,Sjoberg:76,
Kuklov.Svistunov:03,Sellin.Babaev:18}. The total current, is the sum 
of the supercurrents in individual components, which have a similar structure 
as in \Eqref{Eq:Andreev-Bashkin}.  
The first term in \Eqref{Eq:FreeEnergy:2} is responsible for the 
condensation of superconducting electrons such that, in the ground state, 
$\Psi^\dagger\Psi\neq0$. Many two-component superconducting states spontaneously 
break  $\groupU{1}\!\times\!\groupU{1}$, $\groupU{1}\!\times\!\groupZ{2}$ 
or $\groupU{1}$ symmetries (see e.g. \cite{Wang.Cho.ea:16,Wang.Fu:17,
Babaev.Sudbo.ea:04,Babaev.Ashcroft:07,Jones:06}). The corresponding symmetry 
breaking potential terms are collected in $V[\Psi,\Psi^\dagger]$.
For strongly type-II superconductors a good approximation is the constant-density 
(London limit), which is equivalent to $\nu \to \infty $ in 
\Eqref{Eq:FreeEnergy:2}. The results of this paper were verified to hold for 
a wide variety of the symmetry-breaking potential terms, whose detailed 
structure is described in the Methods section. 

In two-component superconductors, the simplest vortices feature a $2\pi$ 
phase winding only in one of the components, e.g. when on a close contour 
surrounding the vortex core $\oint\Grad\varphi_1\!\cdot\!d{\bs\ell}=2\pi$, 
while $\oint\Grad\varphi_2\!\cdot\!d{\bs\ell}=0$. These are called fractional 
vortices (for details, see e.g. \cite{Svistunov.Babaev.ea}).  
Topologically nontrivial knotted vortex loops consist of linked or knotted 
loops of fractional vortices in each component. Like in Kelvin's theory, 
there are infinitely many ways to knot and link such objects. Topological 
considerations imply that knotted vortices are characterized by an integer 
topological index $\Q$ (see e.g. discussions in Refs.~\cite{Faddeev:75,
Faddeev.Niemi:97,Babaev.Faddeev.ea:02,Babaev:02b,Jaykka.Hietarinta.ea:08,
Radu.Volkov:08,Babaev:09}). This index, which is conserved when 
fractional vortices in different components cannot cross each other, 
is defined as (see details in Methods): 
\Equation{Eq:DegreeE}{
\Q = -\frac{1}{12 \pi^2}\int_{\mathbb{R}^3}
	\varepsilon_{ijk}\varepsilon_{abcd}\,\zeta_a
	\frac{\partial \zeta_b}{\partial r_i}
	\frac{\partial \zeta_c}{\partial r_j}
	\frac{\partial \zeta_d}{\partial r_k}
	d{\bf r}\,, 
} 
where ${\bs\zeta}=\left(
\Re\,\psi_1,\Im\,\psi_1,\Re\,\psi_2,\Im\,\psi_2\right)/\sqrt{\Psi^\dagger\Psi}$, 
and $\varepsilon$ is the Levi-Civita symbol. The index $\Q$ is always an 
integer, unless  $\Psi$ has zeros. The situation $\Psi=0$ can appear if 
cores of fractional vortices overlap.
Transient vortex states characterised by such topological index are natural
for two-component superfluids and were experimentally observed
\cite{Lee.Gheorghe.ea:18}. In superfluids, however such vortex knots represent
non-stationary object: the vortex knots are unstable against shrinkage and the 
topological index vanishes when the loops shrink.

\begin{figure*}[!htb]
\hbox to \linewidth{ \hss
\includegraphics[width=0.85\linewidth]{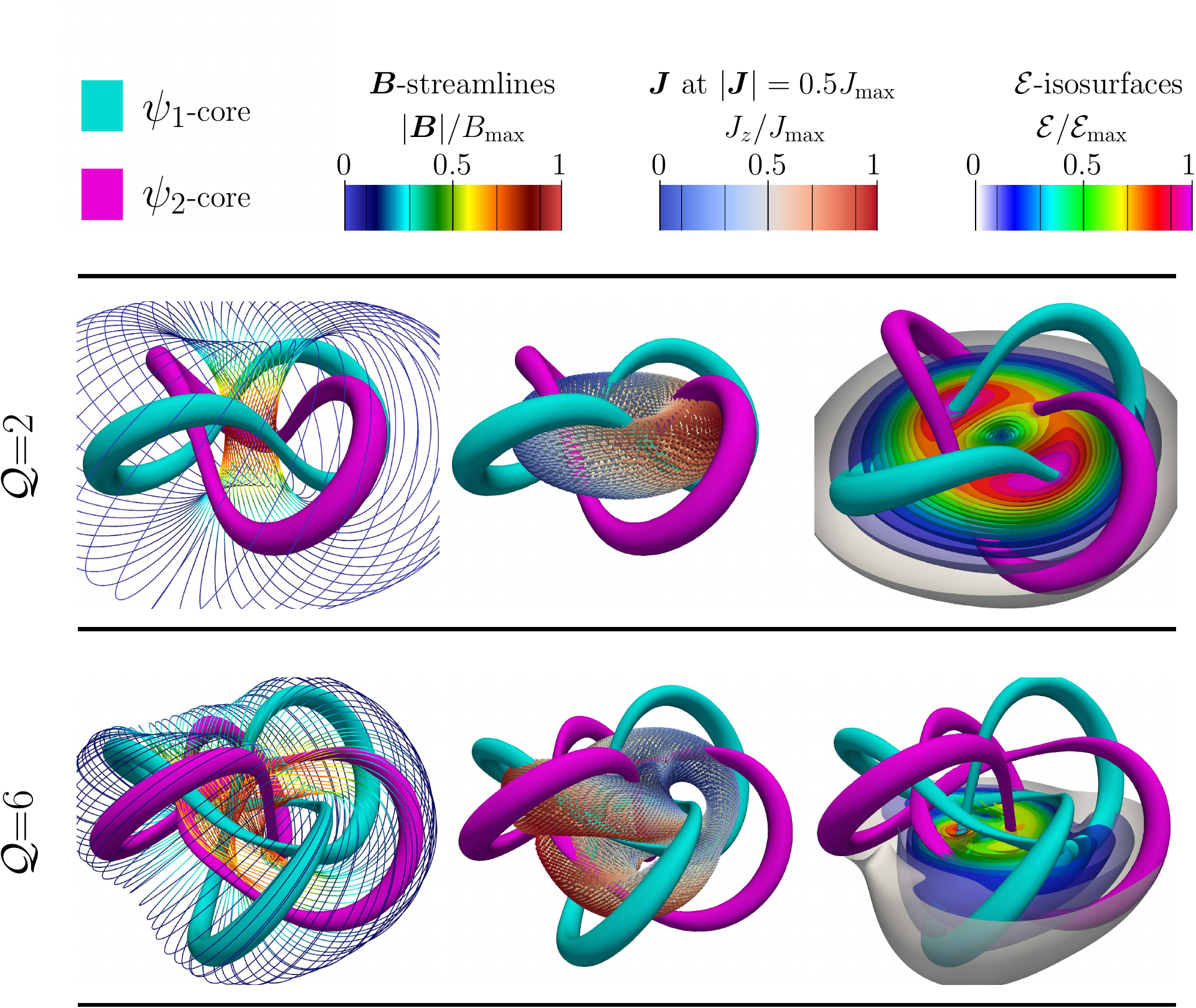}
\hss}
\caption{\small 
Detailed structure of two stable vortex knots with different morphology.
Cyan and magenta tubes denote the positions of the cores of the (fractional) 
vortices defined as the lines where the densities of a given component, 
$|\psi_1|^2$ or $|\psi_2|^2$, vanish. The tubes here are the density-isosurfaces 
such that $|\psi_{1,2}|^2=2.5\times10^{-2}$. The panels on the first column 
show, in addition to the vortex cores, a selection of magnetic field streamlines 
which circulate within the knot (colored according to the magnitude of $|\B|$), 
showing chiral structure of the magnetic field.
The second column displays the total current structure $\J$ on a selected   
isosurface where $|{\bs J}|/J_\mathrm{max}=0.5$. This  also shows their chiral 
structure (the coloring denotes the values of the component $J_z$).
The last panels show isosurfaces of the energy densities of the vortex knots 
solutions.
}
\label{Fig:Knots:1}
\end{figure*}

In order to investigate the existence of stable knotted vortices, 
in two-component superconductors, we performed numerical minimisation of the 
free energy functional \Eqref{Eq:FreeEnergy}, starting from various initial 
states of knotted and linked vortex loops. 
The numerical computations are related, in a way, to the relaxation processes 
of vortex tangles formed due thermal fluctuations or quench. 
Such three dimensional optimisation problem is a highly computationally 
demanding task, which we addressed with a code designed for GPUs 
(see Methods for details).
Upon finding stable knotted solutions for various parameters of the model 
\Eqref{Eq:FreeEnergy}, a detailed investigation of solutions for various 
values of topological index $\Q$ was performed in the London limit where 
$\Psi^\dagger\Psi=1$ (see Methods for details).
We confirm that typically for superconducting models, vortex knots are 
unstable. This agrees with the phenomenology of common superconducting materials, 
where, just like in superfluids, the vortex loops minimize their energy by 
shrinking. However we find that the properties of knotted vorticity become 
principally different, when the Andreev-Bashkin couplings $\hat{\mu}$ are 
substantially larger than usual gradient couplings $\gamma_a$. 
Such a disparity between coefficients occurs near two kinds critical points. 
First is the phase transition to paired phases caused by strong correlations 
\cite{Kuklov.Svistunov:03,Svistunov.Babaev.ea,Sellin.Babaev:18}. There the 
ratio of the stiffnesses of counter- and co-flows of the two components 
vanishes: which implies that the superconductor acquires arbitrarily strong 
Andreev-Bashkin coupling by approaching close enough that critical point 
\cite{Sellin.Babaev:18}.
Second example is the phase transition to Fulde-Ferrel-Larkin-Ovchinnikov, 
where the coefficients $\gamma_a$ change signs (see e.g. 
\cite{Buzdin.Kachkachi:97}), while the Andreev-Bashkin interaction remains 
non-zero. Hence, even systems with relatively weak Andreev-Bashkin interactions 
$\mu_{ab}$ fullfill the above requirements of the disparity of the coefficients, 
sufficiently close to Fulde-Ferrel-Larkin-Ovchinnikov phase transition.
We find that in such regimes, the energy minimization from entangled vorticity
relaxes to stable vortex knots.

\begin{figure*}
\hbox to \linewidth{ \hss
\includegraphics[width=0.95\linewidth]{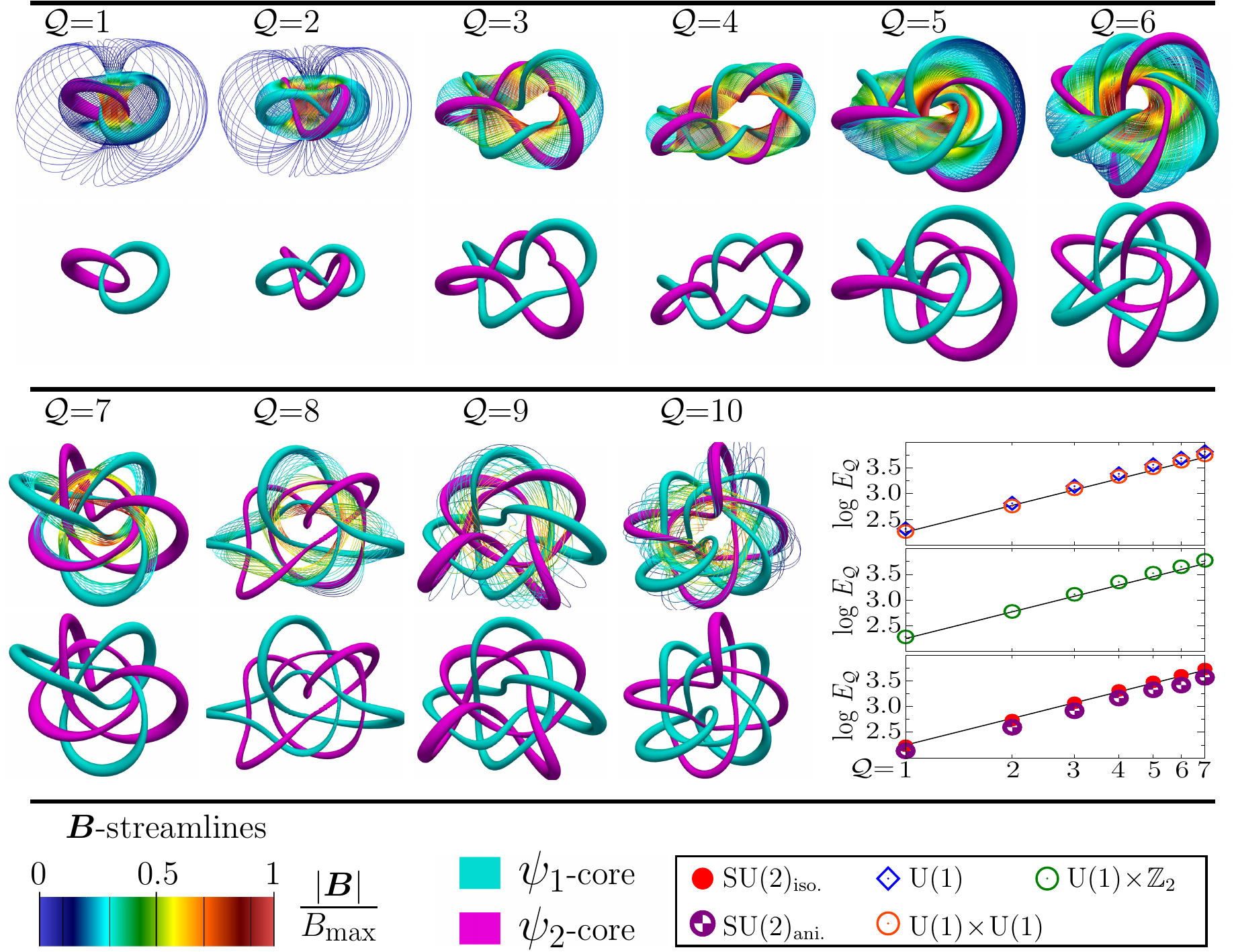}
\hss}
\caption{\small  
Each panel shows stable knots with increasing topological index $\Q$=1-10. 
They consist of two pictures; the lower image displays core structure where, 
as in \Figref{Fig:Knots:1}, cyan and magenta tubes denote the positions of the 
core of the (fractional) vortices in each components. The upper image shows 
in addition to the vortex cores, a selection of magnetic field streamlines 
circulating within the knot. 
Note that when increasing the topological index, the structure of magnetic 
field streamlines becomes increasingly more complex. 
The last panel shows the dependence of the vortex knots energy $E_\Q$, 
as a function the topological index $\Q$.
The curves labelled $\groupSU{2}_\mathrm{iso.}$, $\groupU{1}\!\times\!\groupZ{2}$ , 
$\groupU{1}\!\times\!\groupU{1}$ and $\groupU{1}$ correspond to knotted solutions 
in different models with various symmetry-breaking potentials, while the case 
$\groupSU{2}_\mathrm{ani.}$ corresponds to a symmetry-breaking gradient terms.
}
\label{Fig:Knots:2}
\end{figure*}

The detailed structure for two obtained stable vortex knots is displayed on 
\Figref{Fig:Knots:1}, for a $\groupU{1}\!\times\!\groupU{1}$ superconductor 
where the parameters $\gamma_a=0.02$ and $\mu_{ab}=1$, correspond to a system 
in the vicinity of the above mentioned phase transitions. 
Both topologically different solutions consist of linked and knotted loops
of fractional vortices, which are visualized by the tubes corresponding to 
constant-density-isosurfaces around their cores. 
The energy density of knotted solutions is localized near the knot center, 
thus emphasizing that these objects are particle-like topological solitons 
(i.e. localised lumps of energy). 
The mechanism responsible for the stability of the solutions follows from 
the nontrivial scaling of magnetic field energy, produced by knotted currents. 
During the energy minimisation process that starts from a large vortex tangle, 
the solution first shrinks in order to minimise the kinetic energy of 
supercurrents. This energy gain is eventually counterbalanced by the raise 
of magnetic field energy due to the knotted current configuration. By contrast 
a topologically trivial vortex loop that does not feature helical or knotted 
currents (such as a loop of a single fractional vortex) trivially shrinks to 
zero size.

Akin, to the picture envisaged in Kelvin's theory, the superconducting states 
here support infinitely many stable solutions corresponding to topologically 
different ways to tie vortex knots.
Figure~\ref{Fig:Knots:2} displays ten stable knotted vortex loops, with smallest 
topological indices $\Q=$1-10, in the case of a $\groupU{1}\!\times\!\groupU{1}$ 
superconductor. Animations showing the structure of knotted vortices, and their 
formation can be found in Supplementary Material. We obtained similar stable 
solutions for two-component models that break $\groupSU{2}$, 
$\groupU{1}\!\times\!\groupZ{2}$ and $\groupU{1}$ symmetries. 
The last panel of \Figref{Fig:Knots:2} shows that the energy of knotted 
vortices scales with the topological charge as $E_\Q \propto |\Q|^{3/4}$.
Remarkably the power law is similar to the Vakulenko-Kapitansky law which 
was derived for the solutions of the Skyrme-Faddeev model. 
Thus when increasing the topological index, knotted vortices minimize their 
energy by forming complex bound states of knotted and linked fractional vortices. 
This is again  similar to the picture of matter formation envisaged in Kelvin's 
atom theory.
The solutions with $\Q$=1-4 consist of two linked fractional vortex loops twisted 
around each other different number of times. For $\Q$=5 the solution instead is 
a bound state of two pairs of linked fractional vortex loops. The $\Q$=6 knot 
consists of two linked trefoils knots. For higher topological indices $\Q$=7-10, 
we find that topological structures of the vorticity in different components are 
inequivalent, the solutions thus forming ``isomers". For example the $\Q$=7 knot 
features a fractional vortex in one component forming a trefoil knot, linked with 
two twisted fractional vortex loops of the other component. For each such isomer 
solution there is an energetically equivalent solution where the linked vorticity 
structure is interchanged between the components.


In conclusion, the vortex-atom theory of Kelvin identified chemical elements 
with knotted vortex loop in luminiferous aether. Remarkably, this theory has 
profound similarities, but also some important differences with the physics 
of superconductivity, developed a century later.
In particular, the Meissner effect dictates that superconductors can carry 
magnetic fields and current only in a thin layer near their surface, unless 
they form  quantum vortices. An external magnetic field create vortex lines 
that terminate on superconductor's surface. The field-induced vortices form 
different collective states: lattices, liquids and glasses, all featuring 
distinct transport properties. 
To create a current in the bulk of a superconductor, in the absence 
of an external field, it is required to form a closed vortex loop. Closed loops 
form dynamically: e.g. due to quenches or thermal fluctuations. The crucial 
difference with Kelvin's theory is that in ordinary superconductors vortex loops 
are not energetically stable. This intrinsic instability of a vortex loop 
determines the response of superconductors to an external magnetic 
field, their post-quench relaxation, and their critical properties. 
We demonstrated that under certain conditions the properties of vortex excitations 
in a superconductor change radically and knotted vortex loops become stable, 
akin envisaged in Kelvin theory. We find that this occurs in multicomponent 
superconductors, at least near certain critical points, for example in the 
vicinity of Fulde-Ferrel-Larkin-Ovchinnikov states. The energy associated with 
knotted vorticity starts increasing if vortex knots shrink beyond certain size. 
Moreover the knots form complicated bound states which is strikingly similar to 
the matter formation envisaged in Kelvin theory.
This stability property of vorticity implies radically different
hydro-magnetostatics, compared to  ordinary superconductors. This opens-up 
further questions on the macroscopic properties of these states.

\footnotesize
\vspace{0.5cm}
\noindent\textbf{Online Content} Methods, along with any additional Extended Data display items and Source Data, are available in the online version of the paper; references unique to these sections appear only in the online paper.
\normalsize

\footnotesize
\vspace{0.5cm}
\noindent\textbf{Supplementary Information} 
	is available in the online version of the paper.

\vspace{0.5cm}
\noindent\textbf{Acknowledgements} We acknowledge fruitful discussions 
with Johan~Carlstr\"{o}m, Juha~J\"{a}ykk\"{a}, during various stages of this 
work.
The work was supported by the Swedish Research Council Grants
No.~642-2013-7837, No.~VR2016-06122 and G\"{o}ran Gustafsson Foundation 
for Research in Natural Sciences and Medicine. The computations 
were performed on resources provided by the Swedish National 
Infrastructure for Computing (SNIC) at National Supercomputer 
Center at Link\"{o}ping, Sweden.

\vspace{0.5cm}
\noindent\textbf{Author Contributions }    
F.\,N.\,R.  performed  finite-difference computations. 
J.\,G. performed  finite-element computations. 
All authors contributed to writing the paper.
\normalsize

\newpage
\clearpage
\normalsize

\newpage
\clearpage

\section*{Methods}
\footnotesize

\noindent\textbf{Calculation of topological index of vortex knots.} 
Different vortex knots are characterised by an integer index/invariant 
$\Q$, associated with the topological properties of the maps 
$\groupS{3}\to\groupS{3}_\Psi$.
To calculate that invariant, the superconducting order parameter field 
$\Psi$ is cast into a 4-dimensional vector ${\bs\zeta}=\left(
\Re\,\psi_1,\Im\,\psi_1,\Re\,\psi_2,\Im\,\psi_2\right)/\sqrt{\Psi^\dagger\Psi}$. 
Note that for ${\bs\zeta}$ to be well defined, there should be no zeros of 
$\Psi$, i.e. no overlap of the core centers of fractional vortices in both 
components. 
If cores of fractional vortices in different components can cross each other, 
it generates a point where $\Psi^\dagger\Psi=0$ and the topological index 
$\Q$ is not necessarily an invariant. Namely, it can discretely change from 
different integer values, when fractional vortices in different components 
cross each other.
The non-crossing requirement is always satisfied in the constant total density 
limit, i.e. the London limit ($\Psi^\dagger\Psi=\mathrm{const}$). 
Note that when knots are unstable, the invariant changes when they collapse to a 
size comparable with the numerical lattice size.

Finiteness of the energy implies that the superconductor should be in the 
ground state at spatial infinity. It follows that infinity is identified 
with a single field configuration (up to gauge transformations). Hence, 
the vector field  ${\bs\zeta}(\mathbf{r})$ is a map from the one-point 
compactified space to the target 3-sphere 
${\bs\zeta}:\groupS{3}\,[\cong\mathbb{R}^3\cup\{\infty\}]\to \groupS{3}_\Psi$.
Maps between 3-spheres fall into disjoint homopoty classes, the elements of
the third homotopy group $\pi_3(\groupS{3}_\Psi)$, which is isomorphic to 
integers: $\pi_3(\groupS{3}_\Psi)=\mathbb{Z}$.
Thereby, ${\bs\zeta}$ is associated with an integer number, the degree of 
the map ${\bs\zeta}$: $\mathrm{deg}\,{\bs\zeta}$, which counts how many times 
the target sphere $\groupS{3}_\Psi$ is wrapped, while covering the whole 
$\mathbb{R}^3$ space. 
Field configurations are thus characterized by the topological index 
$\Q:=\mathrm{deg}\,{\bs\zeta}$ which is calculated using equation 
\Eqref{Eq:DegreeE}. 
As discussed, for example, in \cite{Jaykka.Hietarinta.ea:08,Jaykka.Palmu:11}, 
the degree of ${\bs\zeta}$, $\Q$ is equal to the Hopf charge of the combined 
Hopf map $h\circ{\bs\zeta}:\groupS{3}\to \mathbb{S}^2$. 
The formula \Eqref{Eq:DegreeE} was used to calculate the topological 
invariant of the numerically obtained stable knotted vortex configurations. 
It was found numerically to be an integer, with the accuracy of a few percent.

\noindent\textbf{Numerical methods and details of the model.}
The free energy \Eqref{Eq:FreeEnergy} features a potential term 
$U=\nu(\Psi^\dagger\Psi-1)^2$, responsible for the nonzero superconducting 
ground state $\Psi^\dagger\Psi\neq0$. It is supplemented by additional terms, 
which explicitly break global $\groupSU{2}$ symmetry of $U$ down to
different subgroups
\Equation{Eq:Potential}{
V[\Psi,\Psi^\dagger] = \sigma|\psi_1|^2|\psi_2|^2 
+\eta\left(\psi_1\psi_2^*+\psi_1^*\psi_2\right) \,.
}
The solutions were obtained for the parameters of kinetic term $\gamma_a=0.02$ 
and of the current coupling coefficients $\mu_{ab}=1$ for $a,b=1,2$ (except in 
the $\groupSU{2}_\mathrm{ani.}$ case described below). The gauge coupling 
constant, used to parametrize the London penetration depth, was set to $g=1$.
In order to demonstrate that the existence of stable knotted vortices does not 
rely on a specific symmetry of the model, we performed computations for various 
representative symmetry breaking potentials:
 $\bs{\groupU{1}}$: 
	$\sigma=-10^{-4}$, $\eta=-5\cdot 10^{-5}$; $\bs{\groupU{1}\!\times\!\groupZ{2}}$: 
	$\sigma=10^{-4}$, $\eta=0$; $\bs{\groupU{1}\!\times\!\groupU{1}}$: 
	$\sigma=-10^{-4}$, $\eta=0$; $\bs{\groupSU{2},\mathrm{iso.}}$: 
	$\sigma=0$, $\eta=0$;
  $\bs{\groupSU{2},\mathrm{ani.}}$:  
	for this particular case the ground state is invariant under $\groupSU{2}$ 
	rotations, but the current-current coupling coefficients 
	were chosen to   break that symmetry: 
	$\mu_{11}=\mu_{22}=0.97$ and $\mu_{12}=\mu_{21}=1$.
The symmetry-breaking parameters were chosen to be small to avoid important 
changes in the intrinsic length scales, and thus to prevent substantial changes 
of the solution size relative to the numerical lattice spacing. That allowed a 
quantitatively accurate comparison of knotted solutions for different 
symmetry-breaking potentials.

In our preliminary simulations, we considered various values of the coefficient 
$\nu$. When stable vortex knots exits they have no zeroes of total density.
The stability properties of the solutions were typically better 
by increasing $\nu$ (i.e. stronger type-II regimes). However, such a regime 
significantly inhibits the convergence of conventional methods of minimization. 
This phenomenon is directly related with a well-known disadvantage of penalty 
function method.  
Thus after finding stable solutions, with different indices $\Q$, 
in several simulations with variable total density, we reduced the size of 
the parameter space by focusing on the London limit where $\Psi^\dagger\Psi=1$ 
in order to systematically investigate the solutions with topological index 
ranging from $\Q=1$ to $\Q=10$.
\\

For computationally efficient investigation of the London limit we used the 
``Atlas'' method, which is based on an efficient navigation in the coordinate 
charts for the  order parameter manifold, which here is an $\groupS{3}_\Psi$.  
(for details of the method see Supplementary Material in 
Ref.~\cite{Rybakov:2015}). Its advantage is that it automatically satisfies 
the constraint, being in conjunction with conventional unconstrained minimisation 
techniques.

The fields were discretised using a second-order accuracy finite-difference 
scheme, on an homogeneous cuboidal mesh with lattice spacing $0.5$. The grid 
used consist on $160^3$ nodes for the solutions with $Q\leq7$ and $224^3$ nodes, 
for higher topological charges. In order to ensure that the solutions are not 
artefacts of a finite simulation domain, we considered both `fixed' 
and `free' boundary conditions.
The energy was minimised using a nonlinear conjugate gradient (NCG) method with Polak-Ribi\`{e}re-Polyak formula~\cite{PRrule,POLYAK1969}. 
We used a branched approach to standard NCG by separating all degrees of freedom into two sets. One 
associated with gauge degrees of freedom $\mathbf{A}$, and another with the 
superconducting degrees of freedom $\psi_a$. Accordingly, the conventional linear 
search routine for NCG method was replaced by two-dimensional search. 
By separating the numerical degrees of freedom according to their physical nature, 
this approach significantly accelerates the convergence speed of the algorithm.  
The termination criterion  for convergence was chosen according to 
\cite{PracticalOptimization} with a function tolerance $\tau_F\!=\!10^{-10}$. 
The algorithm was parallelized for NVIDIA CUDA-enabled graphics processor units. 
Calculations were performed on a set of two video cards with microprocessors 
GP102-350-K1-A1. In order to achieve maximum performance, most computations were 
realized using single-precision floating-point format (32 bits). To offset the 
truncation errors from the single-precision arithmetic, we used Kahan summation 
algorithm~\cite{Kahan:1965} and a parallel reduction technique  suitable for 
CUDA \cite{ParallelReductions}.

In order to cross-validate our results, we performed few simulations beyond 
the London limit using a different approach based on a finite-element methods, 
and found consistent results. The fields were  discretized within a framework 
provided by the FreeFem++ library \cite{Hecht:12}. The simulations were 
typically performed  on four, two-sockets nodes, with 8-core Intel Xeon E5-2660 
processors.

\noindent\textbf{Initial states for energy-minimization calculations.}  
The energy minimization calculations were seeded with a vorticity-generating 
initial state based on a hedgehog-type texture~\cite{Skyrme:61}
\Equation{Eq:InitialGuess}{
(\zeta^\prime_1,\zeta^\prime_2, \zeta^\prime_3)  = \mathbf{m} \sin(\chi), 
\quad \zeta^\prime_4 =  \cos(\chi)\,,
}
where the unit vector field $\mathbf{m}$ is defined as
$$ m_1 + i m_2 = \sin(\vartheta)e^{-i Q \phi}, \quad m_3 = \cos(\vartheta), $$
and the shape function
$$ \chi = \pi\left( 1 + (r/r_0)^2 e^{(r/r_0)^2} \right)^{-1} \,, $$
where $r$, $\vartheta$, $\phi$ are the spherical coordinates, and $r_0$ is a 
tunable parameter which sets the appropriate scale of the texture. 
For the initial state to satisfy the appropriate behaviour at 
$r\rightarrow\infty$, the vector $\mathbf{\zeta^\prime}$ has to be rotated: 
$\mathbf{\zeta} = R\cdot\mathbf{\zeta^\prime}$. 
The generic rotation matrix, and a particular one, $\tilde{R}$, in the case 
of the potential yielding a $\groupU{1}\!\times\!\groupZ{2}$ symmetry of the 
ground state, are defined 
\Equation{Eq:Rotations}{
R = \frac{1}{2}
\begin{pmatrix}
 1 &  1 & -1 & 1 \\
-1 &  1 &  1 & 1 \\
 1 & -1 &  1 & 1 \\
-1 & -1 & -1 & 1 
\end{pmatrix}\,,~~
\tilde{R}= \frac{1}{\sqrt{2}}
\begin{pmatrix}
 0 &  0 & -1 & 1 \\
 0 &  0 &  1 & 1 \\
 1 & -1 &  0 & 0 \\
-1 & -1 &  0 & 0 
\end{pmatrix} \,.
}
Thus, at the boundaries of simulation domain the superconducting degrees of 
freedom assume ${\bs\zeta}=(1,1,1,1)/2$,  except in the case of the 
$\groupU{1}\!\times\!\groupZ{2}$ symmetry where ${\bs\zeta}=(1,1,0,0)/\sqrt{2}$.
The vector potential is initially  set to be a pure gauge $\mathbf{A}\!=\!0$.

Such initial state generates linked vorticity for which the topological charge 
$Q$ is an input   parameter. 
We observed that for regimes with stable knots the total topological charge 
remains invariant in the process of minimization. 
To obtain solutions with high topological index  $\Q$, we placed two separated 
textures along the main diagonal of computational domain with charges $\Q_{1,2}$ 
such that $\Q_1 + \Q_2 = \Q$. For example, to construct a $\Q$=1 solution, the 
starting configuration was $\Q_1$=-1, $\Q_2$=2; for $\Q$=2 we used $\Q_1=\Q_2=1$, 
etc.
During the minimization process, the two initially separated textures attracted 
and eventually merged into a single one. Such an approach starting with two well 
separated  vorticity-seeding initial states, instead of a single one is very 
efficient because it breaks spatial symmetries, thus minimizing chances of being 
trapped in long-living unstable or weakly metastable states.

\normalsize

\end{document}